# 4π-periodic supercurrent tuned by an axial magnetic flux in topological insulator nanowires

Ralf Fischer[1*], Jordi Picó-Cortés[2,4], Wolfgang Himmler[1], Gloria Platero[2], Milena Grifoni[4], Dmitriy A. Kozlov[3], N. N. Mikhailov[3], Sergey A. Dvoretsky[3], Christoph Strunk[1], Dieter Weiss[1*]

[1] Experimental and Applied Physics, University of Regensburg, D-93040 Regensburg, Germany

[2] Instituto de Ciencia de Materiales de Madrid (CSIC), 28049 Cantoblanco, Madrid, Spain

[3] A. V. Rzhanov Institute of Semiconductor Physics, Novosibirsk 630090, Russia

[4] Institute of Theoretical Physics, University of Regensburg, D-93040 Regensburg, Germany

**Abstract:**

Topological insulator (TI) nanowires in proximity with conventional superconductors have been proposed as a tunable platform to realize topological superconductivity and Majorana zero modes (MZM). The tuning is done using an axial magnetic flux $\phi$ which allows transforming the system from trivial at $\phi = 0$ to topologically nontrivial when half a magnetic flux quantum $\phi_0/2$ threads the wire's cross-section. Here we explore the expected topological transition in TI-wire-based Josephson junctions as a function of magnetic flux by probing the 4π-periodic fraction of the supercurrent, which is considered as an indicator of topological superconductivity. Our data suggest that this 4π-periodic supercurrent is at lower magnetic field largely of trivial origin, but that at magnetic fields above $\sim\phi_0/4$ topological 4π-periodic supercurrents take over.

## I. Introduction

Majorana zero modes (MZM), quasi-particle excitations having non-Abelian statistics, were predicted to form at topological superconductor boundaries [1-5] and might be a key component for fault-tolerant quantum computing [6,7]. Different concepts to search for MZMs have been suggested [8]. Semiconductor nanowires with strong spin-orbit interaction in which conventional s-wave superconductors induce topological superconductivity have been so far the prevailing platform to search for MZM [9-13] (and references therein). These challenging experiments have already provided good signatures for the existence of MZM at the proximitized region of a semiconductor wire. The drawback is that the Fermi-level ($\mu_F$) must be tuned into a material-dependent, narrow gap opened by the Zeeman effect and that, despite great efforts, the MZM could not be proven beyond doubt [14]. Recently, HgTe and InAs quantum wells subjected to an in-plane magnetic field appeared as a new platform to search for MZM [15,16]. Also, in these experiments, the Zeeman effect is used to tune the system into the topological regime, heralded by a zero-bias peak in experiment.

Here, we explore a conceptually different, yet untried scheme to realize topological superconductivity by combining s-wave superconductors and TI-wires [17,18]. The scheme relies on the band structure of a TI-wire. Without a magnetic field, the subband structure of the wire features a gap at $k_x = 0$, with $k_x$ the wave vector along the wire. In this state, sketched in Fig. 1b (left panel), the band structure is trivial because an even number of modes, say along positive $k$, takes part in transport. Note that every band is degenerate in respect of angular momentum (but not w.r.t. spin). This changes by applying a magnetic flux of one-half flux quantum $\phi_0/2 = h/2e$ ($h$ = Planck constant, $e$ = elementary charge) through the wire's cross-section $A$. Now, the Berry phase, acquired by the spin-momentum locked electron spins around the wire's circumference, is canceled by the magnetic flux along the wire axis. Therefore, the gap closes, non-degenerate, perfectly transmitted surface states appear, and the system becomes topologically non-trivial. The reason is that – for any position of $\mu_F$ in the bulk gap of the TI – the number of propagating modes in one-half of the Brillouin zone is odd [17]. In the case of several occupied subbands, as in our experiments, each (angular momentum resolved) band contributes a single MZM. In this case, an even number of MZM will pair up and form ordinary fermions, but an odd number of MZM remains [17].

The subband structure sketched in Fig. 1b has been probed by transport experiments in various material systems[19-25]. In contrast to semiconductor wires, where Zeeman splitting is the dominant effect, the energy range for achieving topological superconductivity is in TI-wires much larger and not limited to the small Zeeman gap since an odd number of modes exists at $\mu_F$ for any position of $\mu_F$. MZM in these systems are predicted to be exceptionally stable and present over a wide range of magnetic flux, $\mu_F$, and disorder [18]. This has been confirmed in recent theoretical work that investigates the potential impact of various mechanisms that control the strength of the superconducting proximity effect in quasi-1D TI systems by calculating the dependence of the topological phase diagram and the induced quasiparticle gap on relevant control parameters [26].

To probe signatures of MZM in our system, we examine the so-called fractional Josephson effect (FJE) where the ac Josephson current should show 4π- instead of 2π-periodicity as a function of the superconductors' phase difference φ [1,2,27]. MZM, which live at the boundaries of systems with induced topological superconductivity have some similarity with Andreev bound states (ABS) in the induced superconducting gap underneath the superconductor contact. In contrast to ABS, having an even number of crossings in the interval [0,2π] of the energy-phase relation ($E - \varphi$, see Fig. 1d), the presence of MZM causes a spectrum with an odd number of crossings [2]. As the Josephson current is proportional to $dE(\varphi)/d\varphi$ it becomes 4π-periodic. In other words, a 2π-periodic supercurrent flows across the junction in case of conventional Andreev reflection where upon reflection a Cooper pair of charge $2e$ enters (leaves) the superconductor [28]. In contrast, the current becomes 4π-periodic if only a single elementary charge is transferred. A convenient way to look for the FJE is by measuring the junction's $I - V$ characteristic under microwave irradiation [10,29,30]. With a microwave field, quantized voltage (Shapiro) steps appear at voltages $V_n = nhf/2e$ with $f$ the microwave frequency and $n$ an integer [31]. In the case of 4π-periodic Josephson currents the odd steps, i.e., $n = 1,3,..$ are partially missing.

The FJE has been probed, e.g., in two-[30] and three-dimensional TIs [29,32,33], Dirac semimetals [34], and semiconductor nanowires[10,35] showing besides 2π also contributions from 4π-periodic Josephson currents. The occurrence of 4π-periodic supercurrents was interpreted as evidence for topological superconductivity. However, missing Shapiro steps have also been found in topologically trivial Josephson junctions[36]. What we contribute new here is probing the topological transition in TI-wire-based Josephson junctions with its tunable subband structure by tracing the 4π-periodic fraction of the supercurrent as a function of magnetic flux. Below we will use axial and perpendicular magnetic fields to disentangle trivial and topological 4π-periodic currents to explore the trivial-topological transition in our system.

## II. Devices, Device Parameters and Experimental Setup

We fabricate TI-wires from strained, 80 nm HgTe films, epitaxially grown on CdTe, using e-beam lithography and wet-chemical etching [25,37,38]. Fig. 1a sketches the device. The $h/e$ periodic conductance oscillations for different $\mu_F$ revealed the topological nature of the surface states in these wires [25]. Wire widths range between $250 - 580$ nm. Without a gate, $\mu_F$ is typically located at the top of the valence band, so that besides Dirac electrons also bulk holes are present, but the tuning between an even and odd number of modes at $\mu_F$ with an external magnetic field is still viable. Superconducting Nb contacts are placed on the surface of HgTe after removing the capping layers by wet-chemical etching. To enhance contact transparency, we clean the HgTe surface by gentle in-situ Ar$^+$-sputtering. Fig. 1c displays the electron micrograph of a completed device. Below we show data from sample $r1$, which has one of the highest Nb contact transparencies, $D \approx 0.58$ (see [39], S2). The data from three other samples studied show consistent results. Due to wet-chemical etching, the wires have a trapezoidal cross-section. In the case of wire $r1$ the top width is 500 nm, the bottom width 580 nm, and the thickness $d = 80$ nm, resulting in an effective rectangular cross-section of $A = 540$ nm $\times$ 80 nm. The circumference of the wires is always shorter than the phase coherence length, which is of the order of several microns [25]. Thus, transport within the TI junction is phase coherent. The separation $L$ between adjacent superconducting contacts is between 150 nm and 170 nm (wire $r1$). For the topological surface states, this implies a Thouless energy of $\varepsilon_{\text{th}}^{(s)} = \hbar v_F^{(s)}/L \approx 1.6$ meV ($v_F^{(s)}$ is the Fermi velocity of the surface states) larger than the induced gap $\Delta^* \sim 0.59$ meV (see [39], S2). Hence our devices are in the short junction limit [40,41]. The samples are cooled down in a dilution refrigerator with a base temperature of 45 mK. An open-ended coaxial cable, placed a few millimeters near the sample, provides a microwave field with frequencies between 1.5 and 12 GHz. The $I-V$ trace with and without applying microwaves is shown in Fig. 1e. Without microwaves a supercurrent of about 1 µA flows across the junction. With microwaves on, Shapiro steps appear in the $I-V$ trace. For $f = 5.9$ GHz even and odd voltage steps appear while at 5.4 GHz the $n = 1$ step is missing and the $n = 3$ one is barely visible. To quantify the quality of the steps we use bar charts, counting the number of data points within a voltage interval around the plateaus. The corresponding plot in Fig. 1f (right panel) clearly shows the missing $n = 1$ plateau. This indicates a 4π-periodic Josephson effect, while the presence of the higher order odd plateaus indicates that conventional 2π-periodic modes are present [42-44].

### III. Shapiro maps at B = 0

To be quantitative we study in detail the frequency and power dependence, which both affect the Shapiro steps. While the prediction of 4π-periodic Josephson currents is based on microscopic Hamiltonians, the signature of $I_{2\pi}\sin(\varphi)$ and $I_{4\pi}\sin(\varphi/2)$ supercurrents on the Shapiro step spectrum depend on the normal resistance and capacitance of the junction. The corresponding resistively (RSJ) or resistively and capacitively shunted junction (RCSJ) models permit to extract the ratio $I_{4\pi}/I_{2\pi}$ from frequency- and power-dependent measurements of the Shapiro spectrum [42,44].

Corresponding data sets in Fig. 2a-c show, color-coded, Shapiro steps measured at three frequencies as a function of power at $B = 0$. Increasing microwave power increases the AC-current flowing across the junction. From the $I - V$ curves, we extract the histograms (bin counts as in Fig. 1f) at fixed power and frequency by dividing the $V$-axis into small intervals of $0.25\ hf/2e$ and counting the data points within them. The color maps show these histograms as a function of microwave power. Yellow regions mark an accumulation of data points and, therefore, a flat line in the $I - V$-traces, i.e. Shapiro steps. Evaluating the maps, we limit ourselves to the low power regime. For higher powers, oscillations appear, which are well described by Bessel-functions [45], but not relevant here. For $f = 6.6$ GHz all Shapiro steps are visible; steps having lower index $n$ appear at lower power. By reducing the frequency to $f = 5.4$ GHz the first steps $n = \pm 1$ becomes completely suppressed and the third ones are slightly reduced. The sequence of all other steps is unchanged. At $f = 3.7$ GHz the third steps are also fully quenched while the fifth ones are strongly reduced. So far, only a missing first step was observed [10,29,34], except in Josephson junctions made of 2D HgTe [30]. The absent higher odd index steps $n > 1$ show the high quality of our samples and prove that hysteresis, occurring on a smaller bias current scale, is not the origin for the missing steps.

We resort first to a heuristic estimation of $I_{4\pi}$ based on Ref. [29]. For that, we plot in Figs. 2d,e the histograms' power dependency of the above Figs. 2a,b at fixed voltages $V_n = nhf/2e$, for $n = 0,1,2$. From these line cuts we extract the maximum step size (bin count) of the first step, $w_1$, and the second one, $w_2$, as shown in the graphs. Fig. 2f shows the ratio $w_1/w_2$ as a function of microwave frequency. The ratio $w_1/w_2$ is nearly constant for frequencies below 5 GHz but rises sharply for higher $f$. The frequency $f_{4\pi} \approx 5.9$ GHz at which $w_1/w_2 = 1$ holds, is a measure of the amplitude of the 4π-periodic supercurrent, given by $I_{4\pi} = \frac{f_{4\pi} h}{2eR_N} \approx 57$ nA, where $R_N$ is the normal-state resistance of 214 Ω [44]. This corresponds to $I_{4\pi}/I_C \approx 6$ %. Observing a 4π-periodic supercurrent at $\phi = 0$, albeit small, is surprising, as the wire subband structure is topologically trivial. It is conceivable that ballistic trivial modes without gap [46] and/or Landau-Zener transitions between trivial ABS with a small gap (see Fig. 1d) mimic 4π-periodicity [47]. Indeed, recent experimental work suggests that a few modes with high transparency can undergo Landau-Zener transitions at $\phi = \pi$ and cause 4π-periodicity [36]. The observation of a 4π-periodic Josephson current is thus a necessary but not sufficient proof of topological superconductivity and the presence of MZM [48]. Below, we show data that trace $I_{4\pi}$ as a function of magnetic flux, from $\phi = 0$ to $\phi \sim 0.6\phi_0$, i.e. in a regime where we expect the system to change from trivial to topological.

That we observe an increase of $I_{4\pi}/I_C$ with ϕ with a maximum close to ϕ ~ 0.5ϕ$_0$ is one of the key results of our present work.

IV. Shapiro maps with axial magnetic flux

In Figs. 3a-c we show an example of the data measured at $B = 30$ mT, aligned along the wire axis and corresponding to ϕ $= BA \approx 0.27$ϕ$_0$. The first Shapiro steps, fully present at 5.9 GHz, are substantially weakened at 5.4 GHz and are fully absent below $f = 3.7$ GHz. So, compared to $B = 0$, the crossover frequency $f_{4\pi}$ is shifted to slightly lower frequencies. We obtain $I_{4\pi} = \frac{f_{4\pi} h}{2 e R_N} \approx 55$ nA, slightly smaller than without $B$. However, as $I_C$ has decreased much more with $B$ to $I_C = 260$ nA, a larger fraction of the supercurrent $I_{4\pi}/I_C \sim 0.21$ is carried by the 4π contribution. The figures 3d-f are calculations within the RCSJ model addressed below.

A detailed overview of the $B$-dependence of $I_C$ and $I_{4\pi}$ of four samples is displayed in Figs. 4a and Fig. 4b. Fig. 4a shows that the critical current $I_C$ drops quickly with increasing ϕ, while $I_{4\pi}$ in Fig. 4b decreases much less. The corresponding ratio $I_{4\pi}/I_C$ is plotted in Fig. 4c, displaying a maximum when about half a flux quantum ϕ$_0$/2 threads the trapezoidal cross-section of the wires. The effect is largest for wire $r1$. For $r1$, $I_{4\pi}/I_C$ increases from about 6% at $B = 0$ to 43% at $B = 55$ mT. Upon further increase of $B$ we expect the perfectly transmitted state to disappear again and $I_{4\pi}/I_C$ to drop. At $B \sim 70$ mT (ϕ/ϕ$_0$~0.63) we observe no missing Shapiro steps anymore down to $f = 1.5$ GHz, which is the lowest frequency at which Shapiro steps get resolved. The blue point of wire $r1$ in Fig. 4c at ϕ/ϕ$_0$~0.63 can be regarded as an upper limit for $I_{4\pi}/I_C$. Assuming that just below $f = 1.5$ GHz odd Shapiro steps start to disappear and assigning $f_{4\pi}$~1.5 GHz we arrive at $I_{4\pi}/I_C$~24 %. Note though that $I_{4\pi}/I_C$ might be significantly lower as is indicated by the large error bar. Besides $r1$, the data of three other samples with different geometries are shown in Fig. 4a-c. The properties of these samples can be found in the Supplemental Material [39], S2. All samples show a similar dependence of $I_{4\pi}/I_C$ on the flux ϕ although the peak height near ϕ/ϕ$_0$~0.5 varies strongly. This variation is most likely linked to the $B$-dependence of the $I_{2\pi}$-periodic current, which dominates the critical current $I_C$. That $I_{4\pi}/I_C(B)$ has a maximum sometimes slightly below ϕ $=$ ϕ$_0$/2 is expected as this, together with the $B$-range in which the perfectly transmitted mode occurs, depends on many experimental parameters, like the exact position of μ$_F$, disorder, or the exact shape of the nanowire [49]. It also depends on the distribution of transparencies between the superconductor and individual transport channels [26].

## V. Modelling the experimental data with the RCSJ-model

So far, we have estimated $I_{4\pi}$, using a heuristic approach based on the RSJ model, which neglects several effects that may influence the observed Shapiro signatures. In addition, we have performed numerical simulations based on an extended capacitively-shunted junction (RCSJ) model aiming to reproduce the experimental results as accurately as possible. Details of the model can be found in the Supplemental Material [39], S3. The results for $B = 30$ mT are presented in Fig. 3d-f (the results for other $B$-values are shown in [39], S6). As shown there, the theoretical model accurately predicts the Shapiro response of the junction as a function of rf power and frequency. Aiming to match the experimental data, the results include an effective capacitance, which we estimate based on the model of Ref. [50] (see [39], S2 and S5). The model also takes excess currents in the samples into account. Excess currents affect the estimation of $I_{4\pi}$ by altering the effective resistance at small voltages as compared to its normal state value. From the measured $I - V$ curves (see Fig. S1a, [39]), we see that the resistance at $V \simeq 0$ ranges from 96 Ω at $B = 0$ to 157 Ω at $B = 50$ mT, which would yield a larger value for $I_{4\pi}$ than estimated. On the other hand, the RSJ model is known to overestimate $I_{4\pi}$ compared to the capacitively shunted model [43]. Moreover, the model includes thermal effects due to Joule heating in the sample by involving the dissipation of the applied electric power via the electron-phonon coupling (see [39], S4). This allows us to rate the temperature dependence of the different parameters. The simulated values are presented in Fig. 4d. Comparing the RSJ and the RCSJ traces in Fig. 4d shows that the value of $I_{4\pi}/I_C$ depends strongly on the underlying model. Both analysis methods, however, show qualitative agreement, i.e. an increase of $I_{4\pi}/I_C$ with increasing flux with a maximum near half flux quantum. Theoretically, the exact position of this maximum (transition) depends on several factors [26] and may occur before $\phi_0/\phi = 0.5$.

## VI. Distinction between trivial and topological 4π-periodic supercurrents

Despite the excellent agreement between theory and experiment and the maxima of $I_{4\pi}/I_C$ around $\phi_0/\phi = 0.5$, we need to distinguish between trivial (e.g., via Landau-Zener transitions) and topological origin of the 4π-periodic supercurrent. To clarify this, we have measured the Shapiro maps in an axial and perpendicular magnetic field as shown in Figs. 5a,d. For perpendicular $B$, the system is trivial and no $I_{4\pi}$-supercurrent of topological origin is expected. The decrease of $I_C$ with $B$ is in both configurations quite similar (see, S7), which allows us to directly compare the Shapiro maps at specific magnetic field values. Corresponding data for parallel (perpendicular) $B$ are shown in the upper (lower) row of

Fig.5. At 20 mT at $f = 5.4$ GHz, the first Shapiro step is suppressed in both configurations. Thus, we expect a similar transition frequency $f_{4\pi}$ and 4π-periodic current $I_{4\pi}$. This suggests that in this low field range $I_{4\pi}$ is largely of trivial origin. At higher $B$ the situation changes. For $B = 40$ mT at $f = 3.7$ GHz and for $B = 45$ mT at $f = 3.1$ GHz, the first Shapiro step is suppressed for parallel configuration (Figs. 5b,c) while fully present if $B$ is oriented perpendicular to the wire. Thus for $B > 30$ mT, $I_{4\pi}$ is consistently lower for perpendicular configuration. These observations can be explained by trivial 4π-periodic contributions, which get suppressed for both configurations with increasing $B$ while the topological perfectly transmitted mode emerges with increasing $B$. This also explains the decrease of $I_{4\pi}/I_C(B)$ in Fig. 4d for $\phi/\phi_0 > 0.4$ as the topological 4π-periodic contribution is expected to vanish again.

## VII. Conclusion

To summarize, our data and simulation suggest that around $\phi_0/\phi \sim 0.5$ the observed 4π-periodic supercurrent is largely of topological origin. Our experiments are hence a first indication that one can indeed switch between trivial and topological superconductivity with an axial magnetic flux in topological insulator wires, as suggested theoretically [17,51,52].

**Acknowledgment**

The work was funded by the European Research Council (ERC) under the European Union's Horizon 2020 research and innovation program (Grant Agreement No. 787515, 253 ProMotion). R.F. obtained funding from the Elite Network of Bavaria (K-NW-2013-258, Doctorate Program "Topological Insulators"). D.A.K. was supported by the Ministry of Science and Higher Education of the Russian Federation, Grant No. 075-15-2020-797 (13.1902.21.0024). N.N.M. and S.A.D. acknowledge support by RFBR Grant 18-29-20053. M.G. acknowledges support from the German Research Foundation (DFG) Project-ID 314695032-CRC 1277 (Subproject B04). G. P. acknowledges the Mercator Fellow position by the DFG via CRC 1277 in Regensburg, as well as Spain's MINECO through Grant No. MAT2017- 86717-P and the CSIC Research Platform PTI-001.

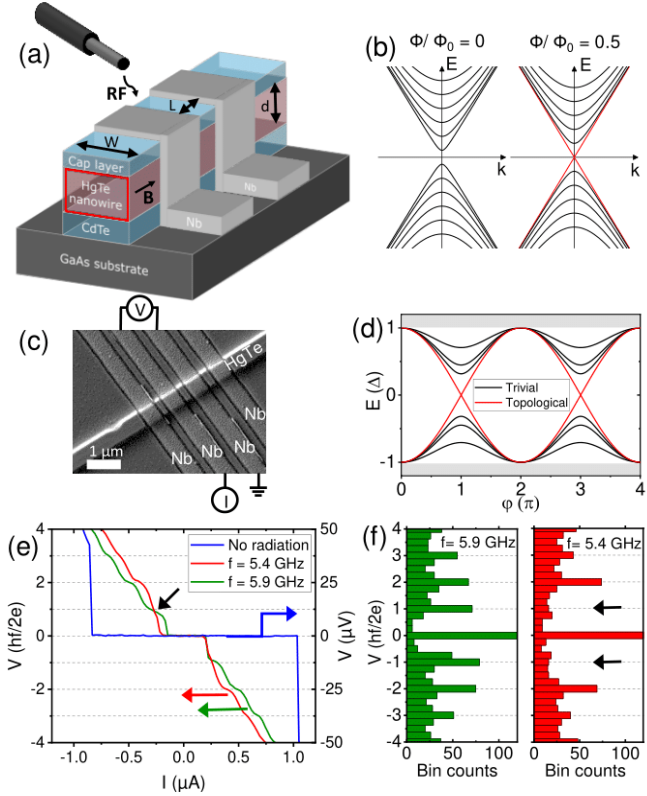

FIG. 1. Sample layout, band structure, and measurement principle. (a) Cartoon of the sample layout showing the HgTe wire and the two Nb contacts which form the Josephson junction. The "spin-less" topological surface states are shown in red. The magnetic field is oriented parallel to the wires' axis. An antenna near the sample radiates microwaves. (b) Subband structure of the surface states of three-dimensional topological insulator nanowires. For zero magnetic flux $\phi = Bwd = 0$ the spectrum is gapped with each band degenerate with respect to angular momentum. When $\phi = \phi_0/2$, a gapless, non-degenerate mode appears. (c) Electron micrograph showing the superconducting Nb stripes placed across the HgTe wire. Only the inner two stripes were used. (d) Spectrum of Andreev bound states with and without MZM. In the trivial case, a gap remains at $\varphi = \pi$, and the energy phase relation is $2\pi$-periodic. In the topological non-trivial case MZM form at $\varphi = \pi$ causing $4\pi$-periodicity and a $4\pi$-periodic Josephson current. (e) I-V trace of nanowire $r1$ (blue trace) at a temperature of 50 mK. The junction has a critical current of $I_C \sim 1$ µA. By adding microwave radiation Shapiro steps appear in the $I-V$ traces (red and green traces). At 5.9 GHz the $n = \pm 1$ steps are visible, while at a lower frequency of 5.4 GHz the first steps are absent (black arrow). (f) Histogram of the data shown in (e).

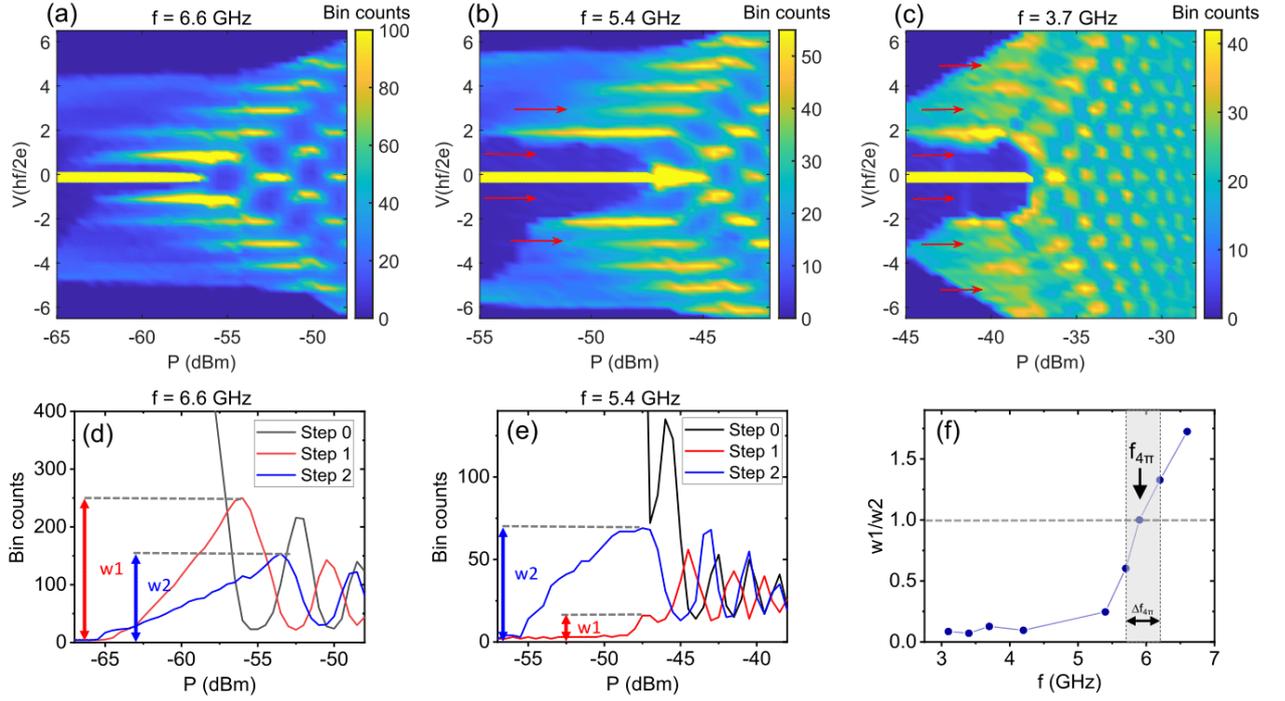

FIG. 2. Frequency dependence of Shapiro steps at $B = 0$. (a-c) Color map of the bin counts of sample $r1$ at frequencies $f = 6.6$ GHz, 5.4 GHz, and 3.7 GHz. The yellow color marks a step in the $I - V$ traces. For the highest frequency, all Shapiro steps are visible. By lowering the frequency, the first step disappears and the third one becomes slightly suppressed. Red arrows mark missing or suppressed steps. For the lowest frequency shown also the third step is strongly suppressed, while the fifth one is further reduced. (d-e) Amplitude of the Shapiro steps as a function of the radiation power for the above color map, respectively. At $f = 6.6$ GHz the maximum amplitude $w1$ of the first Shapiro step is larger than of the second one $w2$. For $f = 5.4$ GHz the amplitude of the first step is almost fully suppressed, and the second step dominates. (f) The ratio $w1/w2$ as a function of the frequency. The ratio decreases by lowering the frequency. At the point $w1/w2 = 1$ the second step starts to dominate. We indicate this point as the transition frequency $f_{4\pi}$. The greyish background indicates the uncertainty $\Delta f_{4\pi}$ in the determination of $f_{4\pi}$.

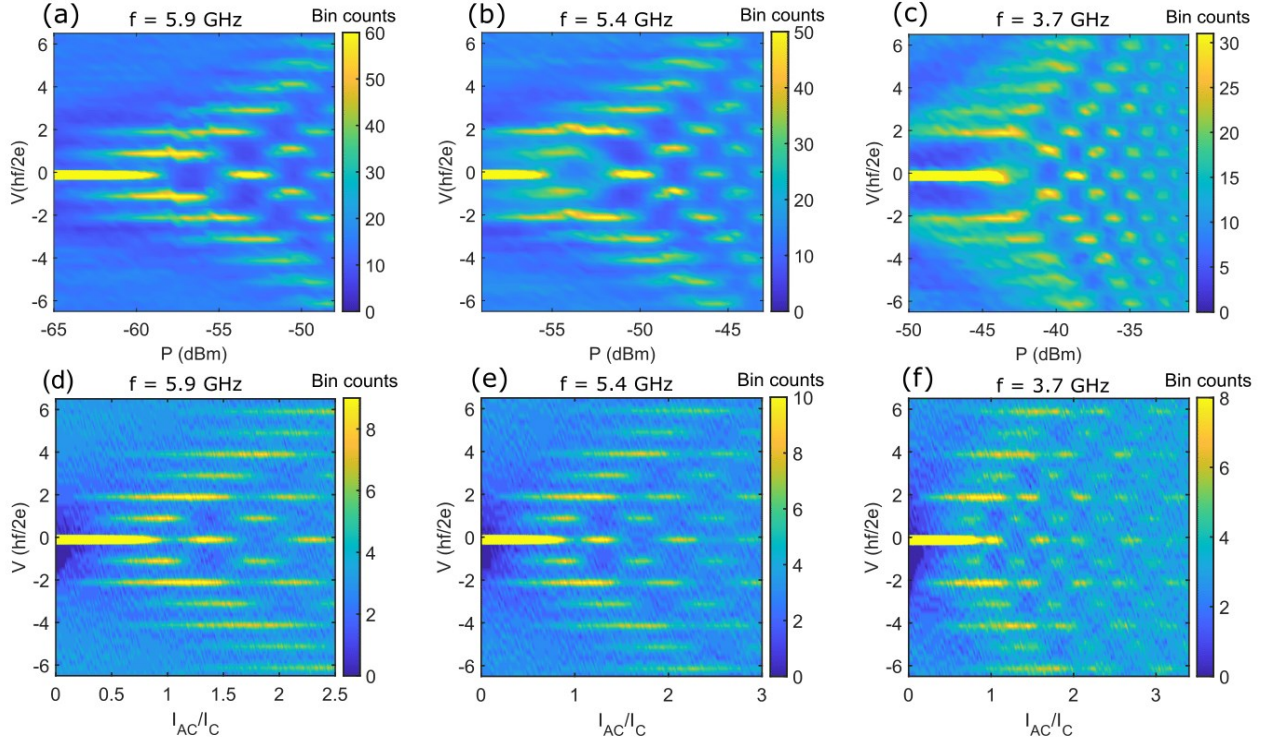

FIG. 3. Frequency dependence of Shapiro steps at $B = 30\ mT$ ($\phi/\phi_0 \approx 0.27$) (a-c) Color map of the bin counts of sample $r1$ at frequencies $f = 5.9$ GHz, 5.4 GHz, and 3.7 GHz. By lowering the frequency, the first step again disappears. However, the transition frequency now slightly changes to $f_{4\pi} = 5.6$ GHz. With $R_N = 216\ \Omega$ we estimate $I_{4\pi}/I_C \approx 21\ \%$ using the RSJ model. (d-f) Numerical simulations using the extended RCSJ-model. The model additionally includes effective capacitance, Joule heating, and a correction due to the excess current, which leads to $R^{\wedge} = 133\ \Omega$ (see Supplemental Material). The model almost perfectly fits the experiment by using $I_{4\pi}/I_C \approx 15.4\%$ for all frequencies. Minor deviations in the $x$-axis scaling of experiment and calculation occur since the exact relation between $P$ and $I_{AC}$ is unknown.

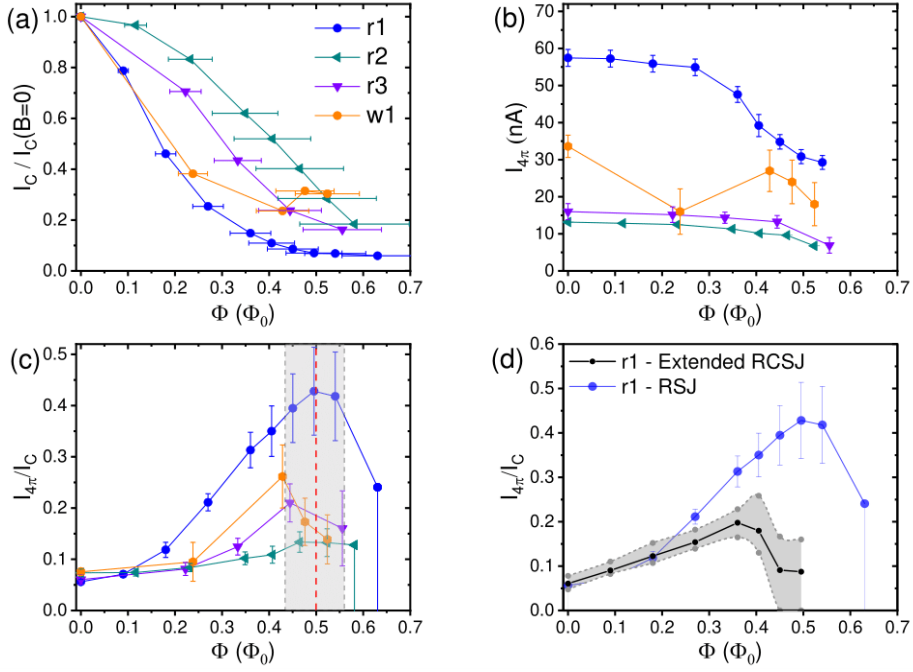

FIG. 4. Magnetic field dependence of the critical currents. (a) The total critical current $I_C$ normalized to its value at $B = 0$ for four wires (see Table S1, Supplemental Material) plotted versus the magnetic field applied in parallel to the wire. The amplitude of $I_C$ is strongly reduced by the magnetic field. (b) The $4\pi$-periodic current $I_{4\pi}$, in contrast, decreases weaker as a function of magnetic flux. (c) The ratio $I_{4\pi}/I_C$ obtained using a heuristic approach based on Ref. [29], plotted as a function of the magnetic field. The red line marks $\phi/\phi_0 = 0.5$ while the greyish background indicates its uncertainty as the cross-sectional area can only be determined within ~12%. All traces display maxima at about $\phi/\phi_0 \sim 0.5$. For wire $r1$ this corresponds to $B \sim 55.5$ mT. (d) Comparison of $I_{4\pi}/I_C$ extracted from wire $r1$ using the heuristic approach used in Fig. c and simulations employing the extended RCSJ-model. The ratio $I_{4\pi}/I_C$ is extracted by comparing the experimental and simulated results. The black dots resemble the values for which the simulations fit the experimental data most accurately, while ratios in the grey area also lead to satisfying results and can be regarded as error bars. The extended RCSJ-model gives somewhat smaller values at high magnetic fields than the heuristic estimation but also shows a distinct maximum, which occurs at a lower field. However, the exact position of the maximum in the blue curve is hard to determine due to the large error bars around $\phi/\phi_0 \sim 0$.

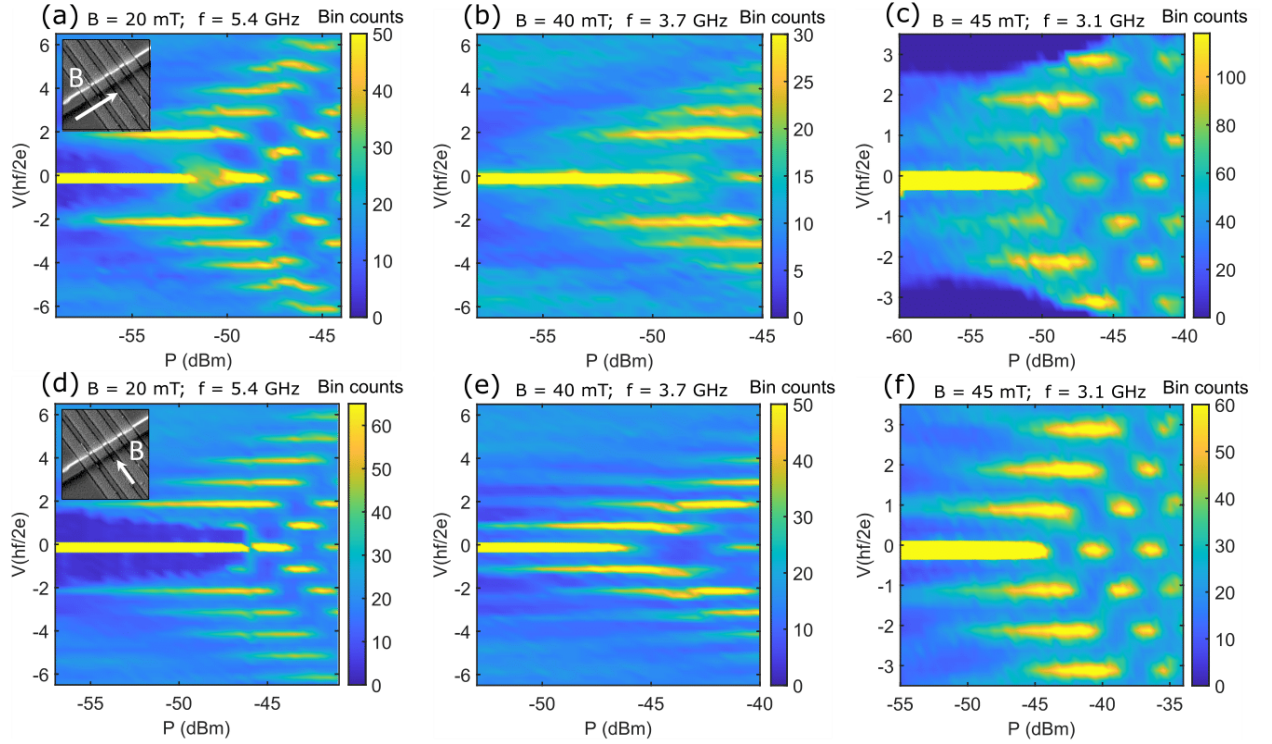

FIG. 5. Shapiro maps at magnetic fields parallel and perpendicular to the wire. (a-c) $B$ parallel to the wire (sample $r1$) as shown in the inset of (a). Color maps of the bin counts at $B = 20$ mT and $f = 5.4$ GHz, $B = 40$ mT and $f = 3.7$ GHz, and $B = 45$ mT and $f = 3.1$ GHz. The first step is strongly quenched or fully absent at these frequencies for the respective magnetic field values. (d-f) $B$ perpendicular to the same wire as shown in the inset of (d). Color map of the bin counts at $B_\perp = 20$ mT and $f = 5.4$ GHz, $B_\perp = 40$ mT and $f = 3.7$ GHz, and $B_\perp = 45$ mT and $f = 3.1$ GHz. While in (d) the first step is quenched similar than in (a), in (e) and (f) all steps are fully visible. The latter is contrary to the observations in (b) and (c). This indicates an additional $4\pi$-periodic component of the supercurrent, which is only present for a magnetic field $B > 20 - 30$ mT, which is oriented along the wire.

## Supplemental Material

### S1. Subband structure of the TI wires

The subband structure of a topological insulator wire with perimeter $P$ is given by $E = \pm \hbar v_F \sqrt{k_x^2 + k_l^2}$ with $k_l = \frac{2\pi}{P}\left(l + \frac{1}{2} - \frac{\phi}{\phi_0}\right)$ [1-3]. Here, $v_F$ is the Fermi velocity of the Dirac surface states, $k_x$ the wave vector along the wire and $k_l$ the transversal wave vector. The latter has the meaning of angular momentum with quantum number $l \in \mathbb{Z}$. The resulting energy spectrum is sketched in Fig. 1(b) of the main part for $\phi = 0$ and $\phi/\phi_0 = 1/2$. In the latter case, the magnetic flux cancels the Berry phase of $1/2$ and non-degenerate, perfectly transmitted states appear, turning the trivial spectrum into a topologically non-trivial one. At $B = 0$, subbands with adjacent angular momentum quantum numbers are separated by $\Delta k_l = 2\pi/P$. With the Fermi wave vector $k_F = \sqrt{4\pi n_s}$ ($n_s$ is the electron density of the surface states) we can estimate the number of occupied subbands as $k_F/\Delta k_l$. With perimeters of the surface states ranging from 560 nm (sample $r2$) to 1280 nm (sample $w1$) and carrier densities $n_s$ between $0.6 \cdot 10^{15}$ m$^{-2}$ (50 nm thick samples) and $1 \cdot 10^{15}$ m$^{-2}$ (80 nm thick samples) the number of occupied subbands of surface electrons ranges between 8 and 22. For estimating the perimeter, we assumed that the surface states are located 5 nm below the HgTe surface.

### S2. Sample fabrication and properties of the fabricated junctions

The HgTe layers (50 - 80 nm) are grown on a GaAs/CdTe substrate by molecular beam epitaxy and are capped by HgCdTe and CdTe layers to avoid $p$-doping via Hg-vacancies and to enhance electron mobility (for details see [4,5]). The HgTe nanowires are patterned by low-energy (3 kV) electron beam lithography (EBL) and a wet-chemical etching solution based on Br$_2$, ethylene glycol, and ultrapure water at 0° C. After aligning the contacts by EBL, the capping layers are removed by wet-chemical etching at the contact area. Before depositing the superconductor, the sample is cleaned by brief Ar+-sputtering (beam voltage: 150 V, duration: ~ 15 s) in an in-situ process. The superconducting contacts consist of 3 nm Ti (e-gun evaporation), 60 nm Nb (sputtering), and 3 nm Pt (e-gun evaporation). We found that the Ti seed layer strongly increases the quality of the interface, while Nb ($T_c \approx 8$ K) defines the properties of the superconducting contacts. The Pt layer prevents Nb from oxidation. The samples are cooled down in a dilution refrigerator with a base temperature of 45 mK. The magnetic field is aligned parallel to the wires' axis, so the magnetic flux through the wire is $\phi = BA$, where $A$ is the cross-sectional area of the wire. An open-ended coaxial cable, placed a few millimeters near the sample, provides a microwave field with frequencies between 1.5-12 GHz. The DC-lines are filtered by $\pi$-filters at room temperature and Ag-epoxy filters [6] as well as RC-filters in the mixing chamber.

From previous experiments [4,5,7,8] on the same or similar wafer material, we can extract the basic properties of our samples. The HgTe films have electron densities of $n_s \approx 1 \cdot 10^{15}$ 1/m² and a mobility of $\mu \approx 20 - 30$ m²/Vs without application of a top gate voltage. This yields a mean free path of $l_{MFP} \approx 1.5 - 2.2$ µm. The Fermi velocity of the surface states is $v_F = \hbar\sqrt{4\pi n_s}/m^* \approx 4.33 \cdot 10^5$ m/s, where the effective mass $m^* = 0.030 \, m_0$ [9].

The niobium leads have a critical temperature of $T_C \approx 8K$, corresponding to an energy gap of $\Delta_{Nb} \approx 1.2$ meV. For the induced gap $\Delta^*$ in HgTe beneath the Nb film, we can estimate an upper and a lower limit. The upper limit is extracted from $I - V$ characteristic, shown for sample $r1$ in Fig S1a. The slope at high bias voltages represents the normal-state conductance $G=1/R_N$, while for lower voltages Andreev reflections increase the slope [10-12]. In Fig. S1b, one sees that $dV/dI$ stays constant and represents the normal resistance $R_N = 214 \, \Omega$ for bias voltages $V > 1.9$ mV. Hence, we estimate $\Delta^* \leq 0.95$ meV as an upper limit. From the high voltage regime $V > 2\Delta^*/e$ in Fig. S1a we can additionally extract the excess current at $B = 0$, $I_{exc} = 2.75$ µA. Assuming perfect transparency $D = 1$ of our junctions we use the relation $I_{exc} = 8/3 \cdot \Delta^*/eR_N$ from [13-15] to derive $\Delta^* \geq 0.22$ meV as a lower limit. Below we use the average $\Delta^* \approx 0.59$ meV of the two limits as a measure of the induced gap. The number of participating channels we can estimate from the normal resistance $R_N$ using $N = N_{ch}^{(s)} + N_{ch}^{(b)} = (R_N \cdot D \cdot 2e^2/h)^{-1}$ with the number of transmitted channels from surface ($N_{ch}^{(s)}$) and bulk ($N_{ch}^{(b)}$), respectively. The number of bulk channels can be obtained from $N_{ch}^{(b)} = N - N_{ch}^{(s)}$, using the number of surface channels from S1. To estimate the transparency $D$ we use the above expression for the number of channels and the established theory for ballistic point contacts [13-15]. Regarding the latter, we use the expression for the excess current [15]:

$$I_{exc} = 2N \frac{e\Delta^*}{h} \frac{D^2}{1-D} \left[ 1 - \frac{D^2}{2(2-D)\sqrt{1-D}} \ln\left( \frac{1+\sqrt{1-D}}{1-\sqrt{1-D}} \right) \right]. \quad (1)$$

Solving both equations using the parameters of sample $r1$, i.e., $R_N = 214 \, \Omega$ and $I_{exc} = 2.75$ µA we obtain $D \approx 0.58$ and $N \approx 103$.

From the length of the junction ($L = 170$ nm) and the Fermi velocity $v_F^{(s)}$ of the surface states, we obtain the Thouless energy $\varepsilon_{th}^{(s)} = \hbar v_F^{(s)}/L \approx 1.6$ meV. Thus, we find $\varepsilon_{th}^{(s)}/\Delta^* \approx 2.7$ and conclude that our junctions are in or close to the short junction limit [13,14]. In the limit $T \rightarrow 0$, the current phase relation (CPR) reads [16]

$$I_{2\pi}(\varphi, T = 0) = \frac{D \sin(\varphi)}{2\sqrt{1 - D\sin^2\left(\frac{\varphi}{2}\right)}} \cdot \frac{Ne\Delta^*}{\hbar} \quad (2)$$

Hence, with $D \approx 0.58$ one expects a surface contribution $I_c^{(s)}(T = 0) \approx 0.35\, N_{ch}^{(s)} e\Delta^*/\hbar \approx 1.03$ µA to the critical current. Moreover, from the ratio of the effective masses of surface and bulk states, we find $v_F^{(B)} \approx 4.6 \cdot 10^4$ m/s and $\varepsilon_{th}^{(B)} = \hbar v_F^{(B)}/L \approx 0.17$ meV for the bulk hole states. Thus $\varepsilon_{th}^{(B)}/\Delta^* \approx 0.29$, corresponding to a long junction with respect to bulk holes. This leads to an suppression (see Fig. 1 in [17]) of the supercurrent contribution of the bulk by a factor 2 and we find $I_c^{(B)} \approx 2.7$ µA leading to an estimated total critical current of $I_c \sim 3.75$ µA, which is more than three times larger than the measured value. A part of the discrepancy can be explained within the heating model below (section S4). Such agreement can be considered satisfactory, as the electromagnetic environment of our junctions is not optimized. The results we obtain for the other samples are summarized in Tab. S1.

| Sample | W [nm] | L [nm] | $d_{HgTe}$ [nm] | $I_C$ [nA] | $R_N$ [Ω] | $I_{exc}$ [nA] | $\Delta^*_{min} - \Delta^*_{max}$ [meV] | $N_{ch}^{(S)}/N_{ch}^{(B)}$ | $D_{average}$ |
|---|---|---|---|---|---|---|---|---|---|
| r1 | 540 | 170 | 80 | 1020 | 214 | 2750 | 0.22 - 0.95 | 20/83 | 0.58 |
| r2 | 250 | 170 | 50 | 180 | 1200 | 580 | 0.26 - 1.05 | 8/9 | 0.61 |
| r3 | 470 | 160 | 50 | 300 | 724 | 694 | 0.19 - 0.85 | 14/17 | 0.57 |
| w1 | 580 | 150 | 80 | 480 | 320 | 1260 | 0.15 - 1.00 | 22/65 | 0.46 |

TAB. S1 Overview of the different samples. Here, $W$ is the width, $L$ is the length of the junction, and $d_{HgTe}$ is the thickness of the HgTe layer. $I_C$ is the critical current of the junction. An upper and lower limit for the induced superconducting gap $\Delta^*$ is estimated from the voltage-dependence of $R_N$ and the excess current $I_{exc}$, respectively. With an average value for $\Delta^*$ and the number of surface $N_{ch}^{(s)}$ and bulk channels $N_{ch}^{(B)}$ the average transmission $D_{average}$ of the junctions can be determined.

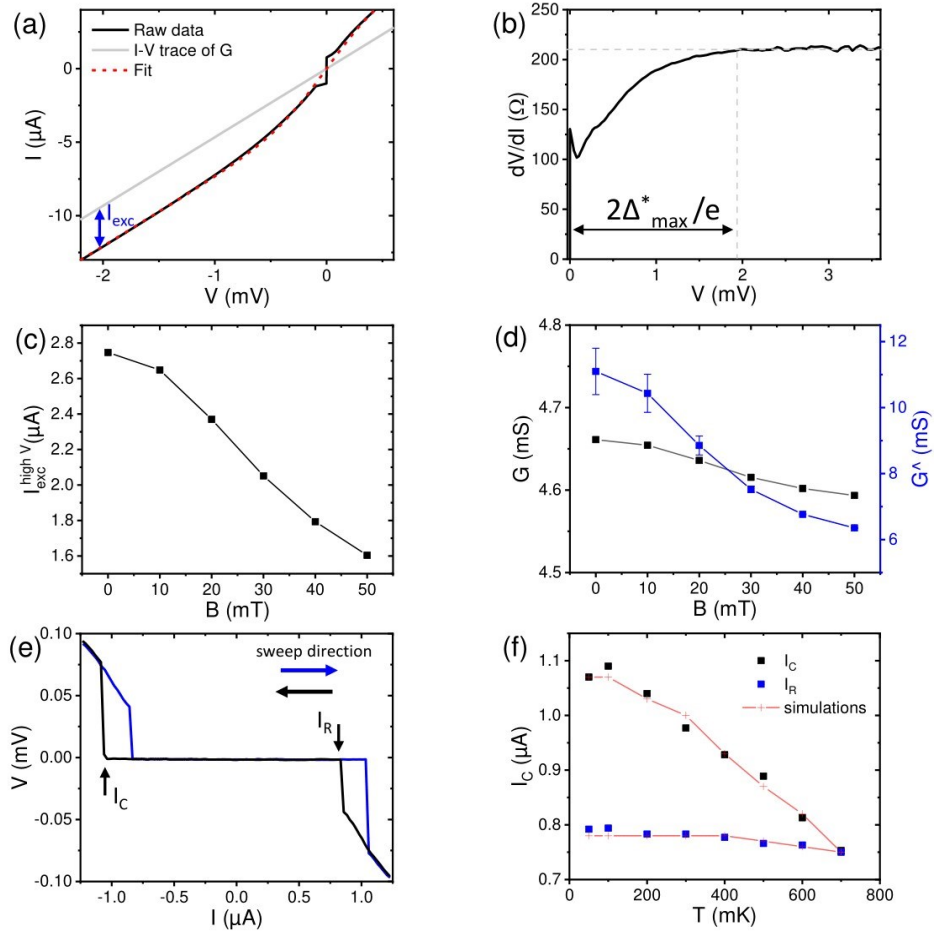

FIG. S1 (a) $I-V$ curve of sample $r1$ at $B = 0$. For high bias voltages, the slope represents the normal-state conductance $G=1/R_N$, while for lower voltages, Andreev reflections influence the trace. The presence of an excess current $I_{exc}^{high\,V} = 2.75$ µA demonstrates the good quality of the interface. The fitting curve $I = VG + I_{exc}\tanh(V/V_{exc})$ perfectly reproduces the trace except for the zero-voltage plateau. (b) The corresponding differential resistance $dV/dI$ plotted as a function of the bias voltage. An upper limit for the induced superconducting gap $\Delta^*$ can be extracted from the curve as the trace starts to deviate from the constant normal-state resistance if $eV < 2\Delta^*$. (c) The excess current $I_{exc}^{high\,V}$ extracted at high bias voltages as a function of the magnetic field parallel to the wire axis. $B = 50$ mT corresponds to $\phi/\phi_0 \approx 0.45$. (d) The black curve shows the normal-state conductance $G$ extracted from the $I-V$ traces at high bias as a function of the magnetic field. The blue points represent the effective conductance $G^\wedge$ at $V \sim 0$, calculated by the parameters $I_{exc}$ and $V_{exc}$. While $G$ changes only slightly, $G^\wedge$ decreases almost by a factor of two between $B = 0$ and $B = 50\,mT$. (e) I-V traces of sample $r1$ at $B = 0$, showing a distinct hysteresis depending on the current sweep direction. (f) The critical current $I_C$ and the retrapping current $I_R$ (a measure of the hysteresis of the $I-V$ characteristic) as a function of $T$ at $B = 0$. The red curve shows the expected values from the extended RCSJ model.

## S3. The extended RCSJ model

The simulations have been carried out employing an extended RCSJ model, which describes the semiclassical voltage response of a current-biased junction. The model includes the effect of excess currents by adding an effective current element in parallel, $I_2$, which introduces a voltage-dependent conductance. Joule heating is also considered by including the explicit temperature dependence of the capacitive and supercurrent elements, as well as its effect on the amplitude of thermal noise.

The extended RCSJ model is given by

$$\frac{\hbar C(T)}{2e}\frac{\partial^2 \varphi}{\partial t^2} + \frac{\hbar G(T)}{2e}\frac{\partial \varphi}{\partial t} + I_{\text{sc}}(\varphi, T) + I_2\left(\frac{\hbar}{2e}\frac{\partial \varphi}{\partial t}\right) \qquad (3)$$
$$= I_0 + I_1 \sin(\Omega_{\text{ac}} t) + I_n(t, T)$$

where $\varphi$ is the phase difference between the two superconductors, $T$ is the quasiparticle temperature, $C$ is the junction capacitance, $G$ is the linear conductance (see below), $I_{\text{sc}}(\varphi, T)$ is the supercurrent, which we assume to have a two-harmonics form

$$I_{\text{sc}}(\varphi, T) = I_{2\pi} f_{2\pi}(\varphi, T) + I_{4\pi} f_{4\pi}(\varphi, T), \qquad (4)$$

where $f_{2\pi}(\varphi, T) = f_{2\pi}(\varphi + 2\pi, T)$ and $f_{4\pi}(\varphi, T) = f_{4\pi}(\varphi + 4\pi, T)$. The two terms in Eq. 4 are the trivial and the $4\pi$-periodic topological contributions, respectively. The functions $f_k(\varphi, T)$ are normalized so that $\max_\varphi\{f_k(\varphi, 0)\} = 1$. $I_0$ and $I_1$ are the dc and ac amplitudes, respectively and $\Omega_{\text{ac}}$ is the ac bias frequency; $I_n(t, T)$ is a Gaussian white noise term with the autocorrelation function $\langle I_n(t_1) I_n(t_2) \rangle = 2k_B T G(T) u(t_1 - t_2)$, with $u(t)$ a unit impulse function, and the term $I_2(V)$ is a correction that accounts for the presence of excess currents, appearing as a result of Andreev reflections in the SN interfaces. Its precise form is given below.

The voltage response of the junction is particularly affected by the $4\pi$-periodic component of the supercurrent when [18]

$$\frac{\hbar \Omega_{\text{ac}}^2 C(T)}{2e}, \frac{\hbar \Omega_{\text{ac}} G(T)}{2e} \lesssim I_{4\pi}. \qquad (5)$$

That is, if the conditions of Eq. 5 hold, when the $4\pi$-periodic part of the current output is comparable to that of capacitive and resistive terms in Eq. 3. It is in this regime where the odd

Shapiro steps are expected to disappear. The second term of Eq. 5 is widely used to estimate the $I_{4\pi}$-contribution in the literature [19-21]. However, it is only valid in the strictly overdamped limit in the absence of excess currents, as predicted by the resistively shunted junction (RSJ) model. The main corrections to this estimation are due to the presence of excess currents and a finite capacitance, which we describe next.

Accounting for the presence of excess currents is particularly important for the estimation of $I_{4\pi}$. The deviation of the current-voltage characteristic from the Ohmic case is seen in Fig. S1a for the case of $B = 0$ mT. In the simulations, we consider an excess current term of the form [22]

$$I_2(V) \simeq I_{exc} \tanh\left(\frac{V}{V_{exc}}\right). \tag{6}$$

Here, $V_{exc}$ is the excess voltage and gives the voltage at which the current switches to the excess branch; $I_{exc}$ is the magnitude of the excess currents i.e. the difference between the measured value of the current at $V \gg V_{exc}$ and the expected ohmic value $GV = I$. Eq. 6 is in excellent agreement with the measured current-voltage curves for the whole range of the magnetic field (see, e.g., Fig. S1a). The excess current is suppressed with increasing magnetic field from $B = 0$ to $B = 50$ mT by a factor 0.57, as shown in Fig. S1c. Hence, we infer that the induced gap $\Delta^*(B)$ is suppressed by the same factor.

Crucially, Eq. 6 produces a voltage-dependent effective conductance. In the limit of high voltages, $V \gg V_{exc}$, the effective conductance is equal to $G$, while for low voltages it is

$$G^{\wedge}(T) = G(T) + \lim_{V \to 0} V^{-1} I_2(V). \tag{7}$$

Shapiro steps appear at voltages $V_n$ corresponding to $\hbar n \Omega_{ac}/2e$ with $n = 0, \pm 1, \pm 2, \ldots$. Hence, for the low frequencies that are employed in the experiment ($\Omega_{ac} \ll 2eV_{exc}/\hbar$) the Shapiro steps appear at $|V| \ll V_{exc}$.

The experimental current-voltage characteristic is fitted by Eq. 6 to obtain approximate values of $I_{exc}$ and $V_{exc}$. Similarly, $G$ is obtained from the slope of the current-voltage curve at high bias, i.e. the excess branches (see Fig. S1a for a representation of the fit). Both $G$ and $G^{\wedge}$ depend on the applied magnetic field, as shown in Fig. S1d. $G^{\wedge}$ has a strong dependence on the magnetic field ( due to the dependence of $I_{exc}$ on the superconducting gap) while $G$ varies only by about 2% between the lowest and the highest magnetic field values.

## S4. Joule heating

The presence of a current bias through the junction increases the effective temperature through Joule heating and leads to thermally-induced hysteresis [23,24]. Following Ref. [25], we include heating by equating the average output power of the junction

$$\langle P(t) \rangle = \langle (I_0 + I_1 \sin(\Omega_{ac} t)) V(t) \rangle, \tag{8}$$

and the power dissipated through electron-phonon coupling [26]

$$\langle P_l(t) \rangle = Q(T^{p+2} - T_l^{p+2}), \tag{9}$$

where $Q$ is a conversion constant, $T_l$ is the lattice temperature and $p$ is the dimensionality of the junction. Comparing both yields the quasiparticle temperature $T$ for each applied $(I_0, I_1)$. The constant $Q$ can be related to the microscopic parameters of the junction through the following model [27]

$$Q = \frac{\beta_l 4 k_B^5 U}{2\pi \hbar^4} \frac{\Gamma(5)\zeta(5)}{v_F u_l^2} \left[ 1 + \frac{16}{\pi^2} \left( \frac{u_l}{u_t} \right)^4 \right], \tag{10}$$

where $U$ is the volume of the central region, $v_F$ is the Fermi velocity, $u_l$ and $u_t$, are the longitudinal and transversal phonon velocities. $\Gamma(n)$ and $\zeta(n)$ are Euler's Gamma and Zeta functions, respectively. From Ref. [28] we obtain $u_l = 2680 \, m/s$ and $u_t = 1620 \, m/s$. The dimensionless constant $\beta_l = k_F^3 v_F^2 m^* / 3 u_l^2 \rho$, where $\rho$ is the mass density of the nanowire, is characteristic of the interaction between the carriers and longitudinal phonons. For the bulk carriers, $\beta_l = 0.001$. From this, we obtain an approximate value of $Q = 7.1 \cdot 10^8$ eV/s K$^5$.

We consider a temperature-independent conductance. The variation of the conductance with temperature at $I_0 \gg I_c$ is taken into consideration by the excess current term, but it may be possible that the conductance varies in the region $I_0 < I_c$ compared to the predictions of the model, since in this area the current-voltage characteristic is flat due to the supercurrent. Regarding the supercurrent, following the results of Section S2, we consider both bulk and surface contributions to the trivial part of the current-phase relation (CPR). Hence, we consider

$$f_{2\pi}(\varphi, T) = \frac{1}{N_{2\pi}} \left( \frac{I_c^{(B)}}{I_c^{(B)} + I_c^{(S)}} f_{2\pi,B}(\varphi, T) + \frac{I_c^{(S)}}{I_c^{(B)} + I_c^{(S)}} f_{2\pi,S}(\varphi, T) \right), \tag{11}$$

where $N_{2\pi}$ is a normalization constant that ensures that $\max_\varphi \{f_{2\pi}(\varphi, 0)\} = 1$.

Regarding the bulk part, we consider the CPR for a long, clean junction, obtained from solving the Eilenberger equations with rigid boundary conditions [16], which is a valid approximation provided that $T < 0.8T_c$. The resulting expression exhibits an exponential decay as the quasiparticle temperature exceeds $T_0 \approx hv_F/4\pi^2 k_B L$.

For the surface contributions, both trivial and topological, we follow Ref. [29] and consider the current-phase relationship corresponding to a driven junction where the voltage response is measured in a time scale below the relaxation time of the Andreev states, occurring due to quasiparticle poisoning and other processes. The CPR is then given by

$$f_{2\pi,S}(\varphi, T) = \frac{1}{N_{2\pi,S}} \frac{\Delta(T)}{\Delta(0)} \frac{\sin(\varphi)}{\sqrt{1 - D\sin^2\left(\frac{\varphi}{2}\right)}} \tanh\left(\frac{\Delta(T)}{2k_B T}\right), \qquad (12)$$

$$f_{4\pi}(\varphi, T) = \frac{\Delta(T)}{\Delta(0)} \sin\left(\frac{\varphi}{2}\right) \tanh\left(\frac{\Delta(T)\sqrt{D}}{2k_B T}\right), \qquad (13)$$

for the trivial and the topological contributions respectively, where $D$ is the transmission of the superconductor/HgTe interface and $N_{2\pi,S}$ is a normalization constant that ensures that $\max_\varphi\{f_{2\pi,S}(\varphi, 0)\} = 1$. If the Andreev states have time to relax to the ground state, the current-phase relationship for the topological component will also be $2\pi-$periodic and the even-odd effect in the Shapiro step profile will not be visible [29]. It is also assumed that the occupation of the Andreev bound states is not modified by non-adiabatic transitions between the bound states.

To match the hysteresis pattern, we consider in the simulations $Q_{\text{sim}} = 3Q$ and $T_{0,\text{sim}} = 3T_0$. This adjustment is warranted when estimating the strength of Joule heating in nanometric junctions due to the inverse proximity effect and the conductance between the normal region and the unproximitized parts of the leads [25,30]. The resulting temperature dependence of the critical and retrapping currents closely matches the measured values (see Fig.S1 f).

## S5. Estimation of the intrinsic capacitance

The capacitance term appearing in the RCSJ model of Eq. 3 accounts for both the geometric capacitance of the junction and the intrinsic capacitance [31] arising from the finite decay time of the Cooper pairs in the normal region. We follow Ref. [32] to estimate the value of the intrinsic capacitance. We consider two parallel contributions arising due to surface and bulk conduction and neglect any phase dependence. This yields an intrinsic capacitance of $C \approx$ 13 fF at $B = 0$. This corresponds to a slightly underdamped junction, according to the value of the Stewart-McCumber parameter $\beta = 2eI_cC/\hbar\,(G^\wedge)^2 \approx 0.41$. This value increases with larger dc and ac biases because of both heating and excess currents but is reduced for higher magnetic fields as the critical current decreases while both the intrinsic capacitance and the conductance remain mostly constant.

## S6. Comparison between theoretical simulations and experiment

Figs. S2-S7 show color maps of the Shapiro step profiles as a function of the power for three different driving frequencies and magnetic field strengths ranging from B = 0 to 55 mT. We present side-by-side comparisons between the simulations obtained from the extended RCSJ model and the measurements on sample $r1$. Note that in the simulations the Shapiro profiles are represented in terms of the amplitude of the ac bias $I_1$, while the experimental results are represented in terms of the rf power $P \propto I_1^2$. For Fig. S2, the experimental results are shown in the main text. The corresponding estimated values of $I_{4\pi}$ are presented in Fig. 4d of the main text, showing a marked increase as the magnetic field approaches half a magnetic flux quantum. Compared to the estimation based on the experiments, the peak in $I_{4\pi}$ in the simulations occurs for lower magnetic fields ($B = 40 - 45$ mT). We choose a value of $I_{4\pi}$ that consistently matches the Shapiro step profiles across the different values of the frequency. Overall, the first step is less quenched at high frequencies in the experiment compared to the simulations, particularly for low magnetic fields. Thermally activated quasiparticle poisoning [30,33,34] or non-adiabatic transitions between the Andreev bound states and the quasi-continuum [35,36] could cause this reduced quenching. Likewise, frequency-dependent damping being stronger at higher frequencies (precisely as discussed in Ref. [37]) could explain this feature, since stronger damping masks the presence of $4\pi$-

periodic modes. For this reason, when estimating $I_{4\pi}$ we prioritize reproducing the lowest frequency results when a single match for all frequencies is not possible. Particularly, for $B = 0 - 30$ mT (see Figs. S2-S4, and Fig.3 of the main text), the results show a Shapiro step profile consistent with stronger damping for the highest frequencies considered, suggesting that the damping increases with frequency. Alternatively, this difference could originate from the temperature dependence of the conductance at low bias, which is not captured by the model employed for the excess current. As an example, in Fig. S8 we have represented a set of simulated Shapiro profiles for $B = 30$ mT and three additional values of $I_{4\pi}/I_{2\pi} = 0\%$, 10% and 30% (corresponding to $I_{4\pi}/I_c = 0\%$, 9.1% and 23%, respectively). For comparison, the value employed in Fig. 3 of the main text is $I_{4\pi}/I_{2\pi} = 18\%$ ($I_{4\pi}/I_c = 15.4\%$). The case $I_{4\pi}/I_{2\pi} = 0\%$ (first row) corresponds to the case without 4π-periodic modes. As can be seen, all steps have approximately the same size, with the first step slightly reduced for $V < 0$ due to hysteresis. For $I_{4\pi}/I_{2\pi} = 10\%$ (second row) we see how the first step is reduced for $f = 3.7$ GHz, but not as significantly as in the experiments. However, the results for $f = 5.9$ GHz are similar to the ones obtained experimentally, with a very slight quenching of the odd Shapiro steps. Hence, we expect that at high frequencies, either a stronger damping or quasiparticle dynamics is masking the presence of the 4π-periodic mode. Finally, for $I_{4\pi}/I_{2\pi} = 30\%$ we observe a completely quenched first step at $f = 3.7$ GHz, in line with the experimental results. However, the higher-order odd steps are also reduced, even at higher frequencies, and the step profile at higher values of $I_{ac}$ is markedly distinct compared with the experiments. For these reasons, we choose a value of $I_{4\pi}/I_{2\pi}$ that is smaller than 30%, (15.4%) for the comparison with the experiment.

For higher magnetic fields (Figs S6-S7), the critical current is significantly reduced and the resistive and capacitive terms in Eq. 3 are of higher importance. In this regime, the effect of the $4\pi$ −periodic supercurrent is harder to observe even at low frequencies. Compared to lower magnetic fields, much lower frequencies are required to access the non-linear regime where the supercurrent dominates the voltage response of the junction. This is reflected in the observed and simulated curves, which show that the first step is less quenched compared to the results for low magnetic field. In this regime, the estimation of $I_{4\pi}$ is less precise. We have represented in Fig. S9 the Shapiro step profile for 55 mT and different values of $I_{4\pi}$ for $f = 3.1$ GHz. As can be seen, the effect of varying $I_{4\pi}$ is hard to distinguish. From the results at higher magnetic fields, it seems that the suppression of the first step is more pronounced at lower frequencies. Due to this, it could be that the values of $I_{4\pi}$ are underestimated at higher magnetic fields because lower frequencies cannot be reached.

The simulations reproduce in large part the observed difference in response between the odd steps, which cannot be explained within the zero-capacitance limit [38]. This reduced quenching of the third and subsequent steps is a consistent feature appearing in several experiments [19-21,30,39].

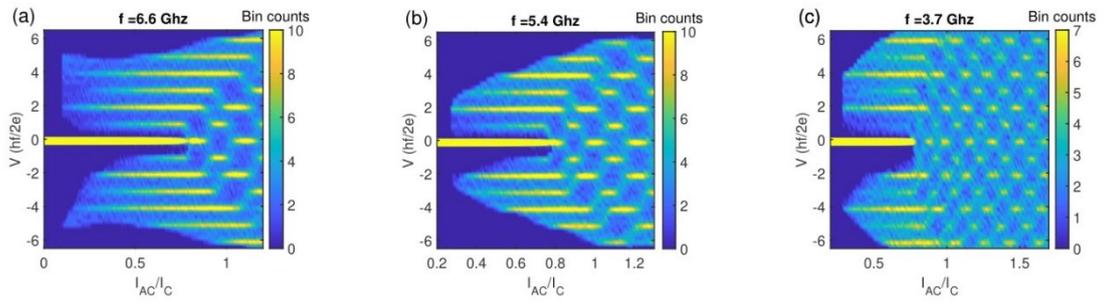

FIG. S2 Frequency dependence of Shapiro steps at $B = 0$ mT ($\phi/\phi_0 = 0$) a-c, Numerical simulations using the extended RCSJ-model. The corresponding experimental data are shown in Fig. 2 (a-c) of the main text. The simulations match the experimental results best for $I_{4\pi}/I_C \approx 6.1\%$.

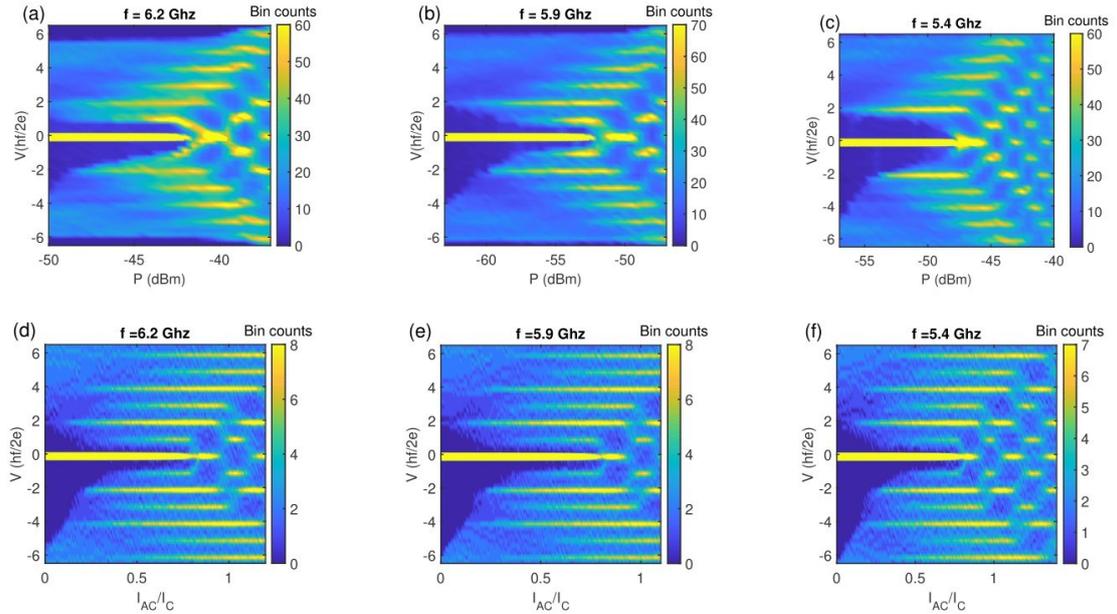

FIG. S3 Frequency dependence of Shapiro steps at $B = 10$ mT ($\phi/\phi_0 \approx 0.09$) (a-c) Color map of the bin counts of sample $r1$ at frequencies $f = 6.2$ GHz, 5.9 GHz, and 5.4 GHz. The transition frequency is

$f_{4\pi} = 5.9$ GHz. With $R_N = 215\ \Omega$ ($G = 4.65$ mS) we estimate from Eq. 3 (the term on the right-hand side with $\Omega_{ac} = 2\pi f_{4\pi}$) $I_{4\pi} = 57.2$ nA, using the RSJ-model. This yields a contribution $I_{4\pi}/I_C \approx 7.1\%$. (d-f) Numerical simulations using the extended RCSJ-model. The simulations reproduce the experimental results most accurately for $I_{4\pi}/I_C \approx 9.0\%$.

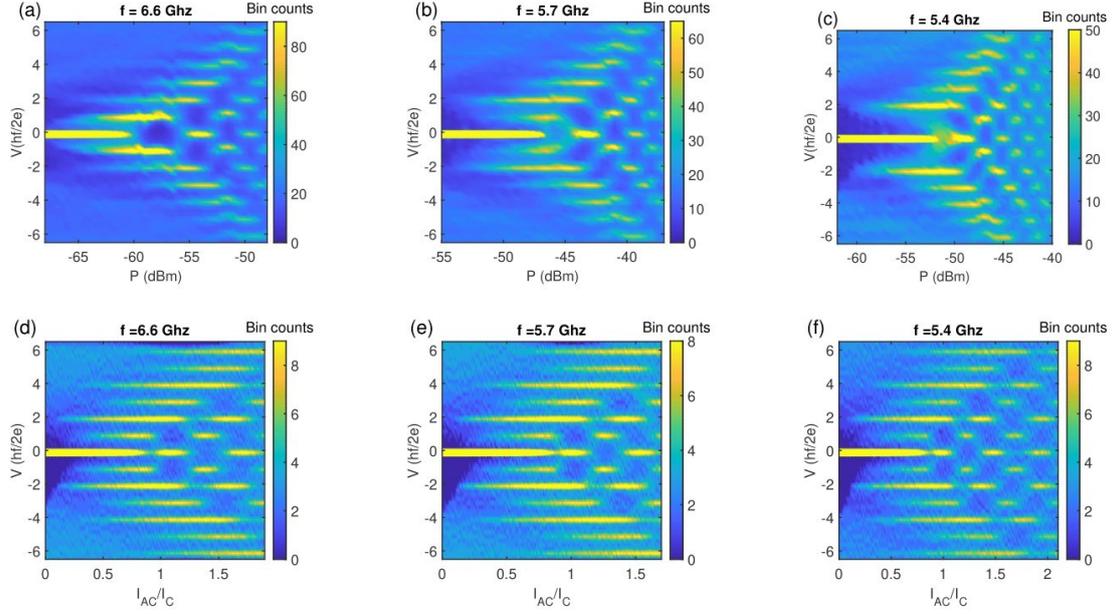

FIG. S4 Frequency dependence of Shapiro steps at $B = 20$ mT ($\phi/\phi_0 \approx 0.18$) (a-c) Color map of the bin counts of sample $r1$ at frequencies $f = 6.6$ GHz, 5.7 GHz, and 5.4 GHz. The transition frequency is $f_{4\pi} = 5.7$ GHz. With $R_N = 216\ \Omega$ ($G = 4.64$ mS) we estimate $I_{4\pi}/I_C \approx 12.2\%$ using the RSJ-model. (d-f) Numerical simulations using the extended RCSJ-model. The simulations reproduce the experimental results most accurately for $I_{4\pi}/I_C \approx 9.2\%$.

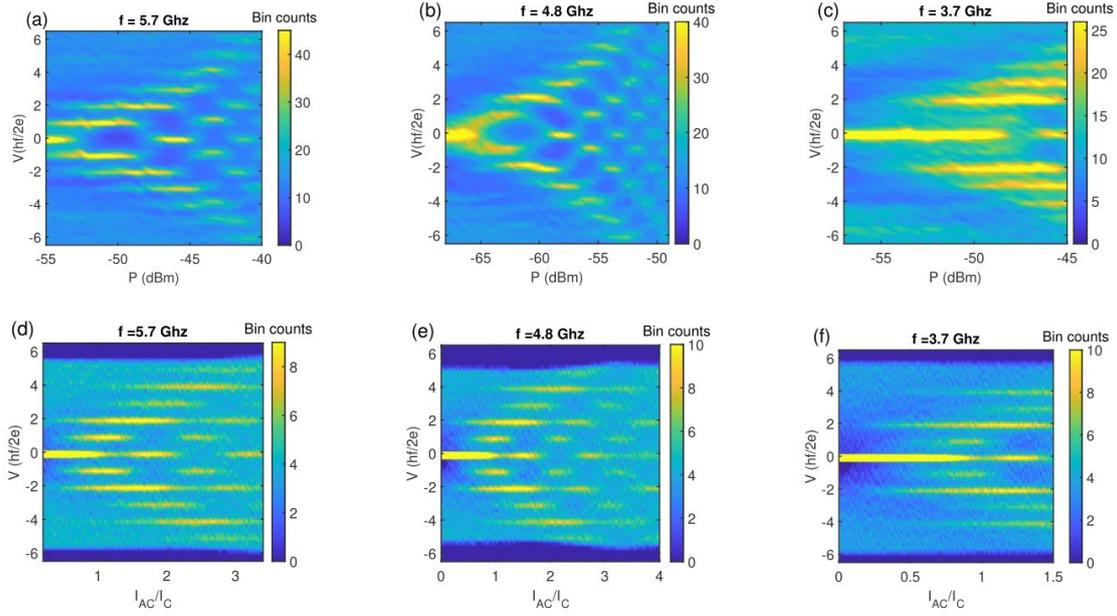

FIG. S5 Frequency dependence of Shapiro steps at $B = 40\ mT$ ($\phi/\phi_0 \approx 0.36$) (a-c) Color map of the bin counts of sample $r1$ at frequencies $f = 5.7$ GHz, 4.8 GHz, and 3.7 GHz. The transition frequency is $f_{4\pi} = 4.9$ GHz. With $R_N = 217\ \Omega$ ($G = 4.61$ mS) we estimate $I_{4\pi}/I_C \approx 31\%$ using the RSJ-model. (d-f) Numerical simulations using the extended RCSJ-model. The simulations reproduce the experimental results most accurately for $I_{4\pi}/I_C \approx 19.8\%$.

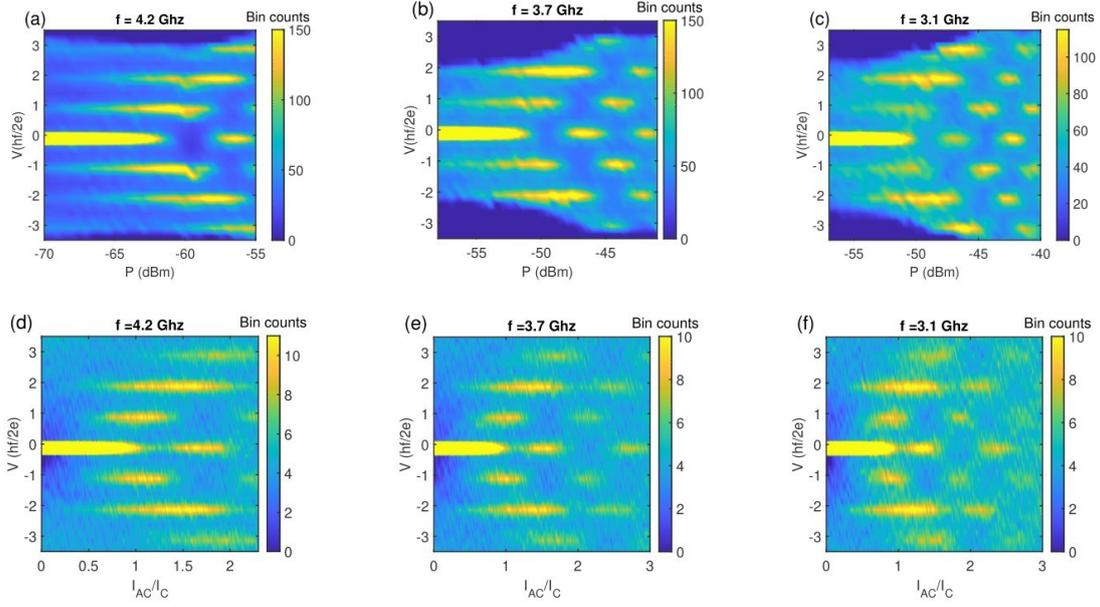

FIG. S6 Frequency dependence of Shapiro steps at $B = 45$ mT ($\phi/\phi_0 \approx 0.41$) (a-c) Color map of the bin counts of sample $r1$ at frequencies $f = 4.2$ GHz, 3.7 GHz, and 3.1 GHz. The transition frequency is $f_{4\pi} = 4.0$ GHz. With $R_N = 217$ Ω ($G = 4.60$ mS) we estimate $I_{4\pi}/I_C \approx 35\%$ using the RSJ-model. (d-f) Numerical simulations using the extended RCSJ-model. The simulations reproduce the experimental results most accurately for $I_{4\pi}/I_C \approx 18\%$.

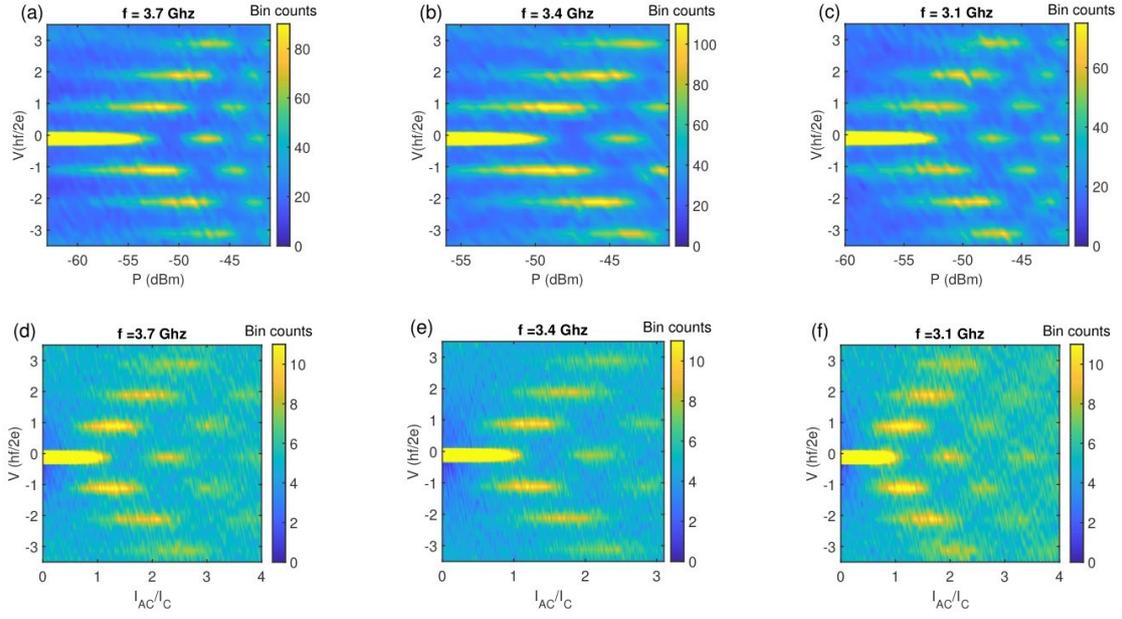

FIG. S7 Frequency dependence of Shapiro steps at $B = 55$ mT ($\phi/\phi_0 \approx 0.5$). (a-c) color map of the bit counts of sample $r1$ at frequencies $f = 3.7$ GHz, 3.4 GHz, and 3.1 GHz, (d-f) Numerical simulations using the extended RCSJ-model. The simulations reproduce the experimental results most accurately for $I_{4\pi}/I_C \approx 8.7\%$.

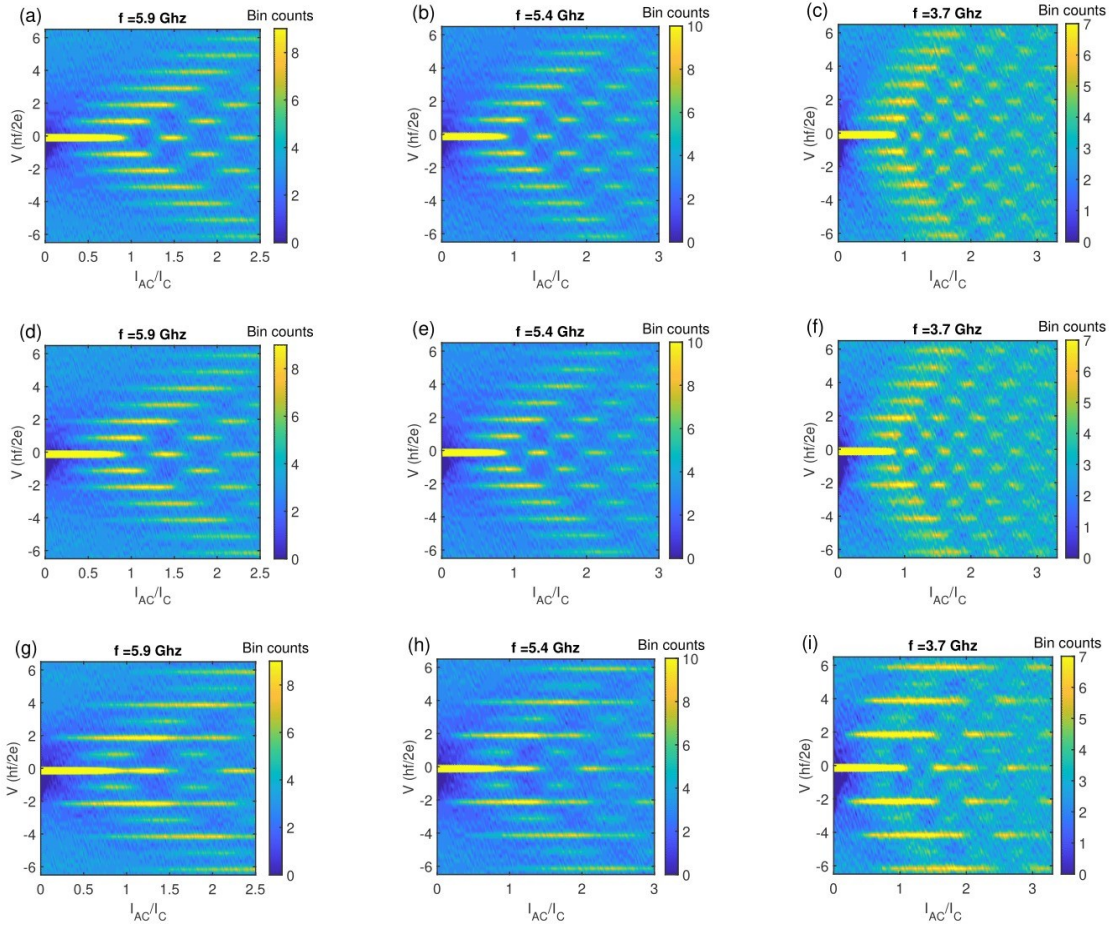

FIG. S8 Simulations at $B = 30$ mT ($\phi/\phi_0 \approx 0.27$) for different values of $I_{4\pi}/I_C$. (a-c) $I_{4\pi}/I_C = 0$ (d-f) $I_{4\pi}/I_C \approx 9.1\%$ and (g-i) $I_{4\pi}/I_C \approx 23\%$

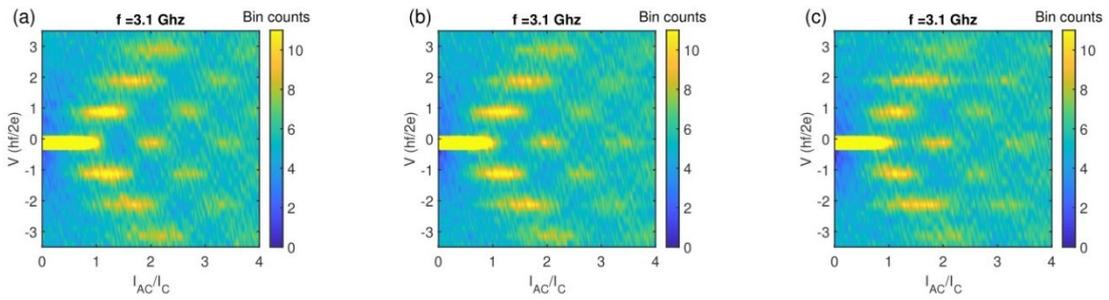

FIG. S9 Additional simulations for $B=55$mT ($\phi/\phi_0 \approx 0.5$) and $f = 3.1$ GHz for different values of $I_{4\pi}/I_C$. (a) $I_{4\pi}/I_C = 0$, (b) $I_{4\pi}/I_C = 8.7\%$ (c) $I_{4\pi}/I_C = 17.4\%$. (Compare with Fig. S7).

## S7. Shapiro maps in perpendicular magnetic fields

In this section, we show supplementary data where the $B$-field is oriented perpendicular to the wire as is sketched in the blue inset of Fig. S10a. For this configuration, no transition from trivial to topologically non-trivial is expected. In Fig. S10a, the critical current $I_C$ as a function of $B_\perp$ (blue trace) is compared to $I_C$ in a parallel field $B_\parallel$ (black trace, same data as in figure 4a of the main text). We plot them as normalized values since the data were taken at different temperature cycles leading to small differences of 7% to 10% for $I_C$. For both orientations of the magnetic field $I_C$ decays quite similarly. Figure S10b and c show the color maps for $B_\perp = 10$ mT at frequencies $f = 5.9$ GHz and $f = 5.4$ GHz. For $f = 5.9$ GHz, we observe a reduction of the first Shapiro step, while it is almost fully absent at $f = 5.4$ GHz. The data for the same magnetic field strength, but parallel orientation, can be found in Fig. S3b, c. As the suppression of the first step is similar for $B_\perp$ and $B_\parallel$, also $I_{4\pi}$ has a similar value. The same we find for $B = 20$ mT (shown in Fig. 5 of the main text) suggesting that the $I_{4\pi}$-periodic supercurrent up to about $B = 20$ mT is mainly caused by trivial effects. The data for $B_\perp = 30$ mT at a frequency $f = 3.7$ GHz is presented in Fig. S10d. Here, we do not observe a clear reduction of the first Shapiro steps, while by applying a parallel field the first step is already fully suppressed at the same frequency (see Fig. 3c of the main text). For $B_\perp = 30\ mT$, we must reduce the frequency to $f = 3.1$ GHz to significantly reduce the first step. The associated color map is shown in figure S10e. This implies that $I_{4\pi}$ for $B_\perp = 30\ mT$ is lower than for $B_\parallel = 30$ mT. For $B_\perp = 40$ mT and $B_\perp = 45$ mT we also do not observe missing or reduced steps at frequencies, where the first step is already quenched for parallel orientation of the magnetic field (see main text). This indicates that there is an additional $4\pi$-periodic contribution, which only arises if a parallel magnetic field $B_\parallel > 20 - 30$ mT is applied. This component we ascribe to tuning the band structure from trivial to topological as expected theoretically. The behavior of $I_{4\pi}(B_\parallel)$ is summarized in Fig. S10f. Due to limited resolution at low frequencies (and low $I_C$) we were not able to probe at lower transition frequencies $f_{4\pi}$ for $B_\perp = 40$ mT and $B_\perp = 45$ mT. Hence, we cannot determine the exact amplitudes of trivial and expected topological component for $B_\parallel > 30\ mT$. The dependence of $I_{4\pi}/I_C(B_\parallel)$ (see Fig. 4d of the main text), which shows a distinct maximum around $B_\parallel = \phi/\phi_0$, indeed suggest that the topological $4\pi$-periodic current dominates at higher magnetic fields.

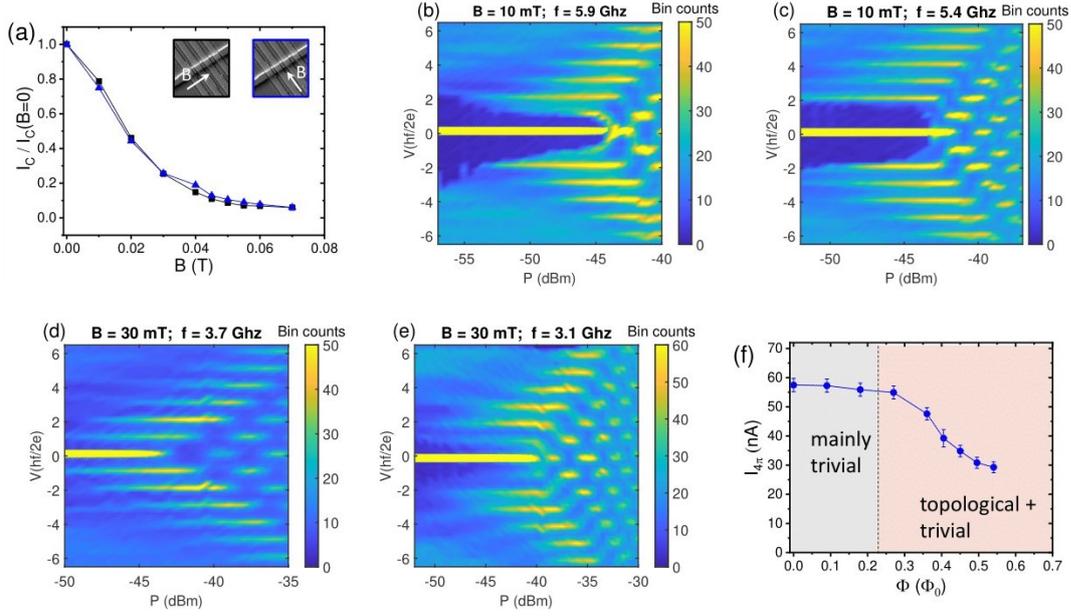

FIG. S10 Data for magnetic fields applied perpendicular to the wire. (a) The total critical current $I_C$ normalized to its value at $B = 0$ plotted for the magnetic field applied parallel (black) and perpendicular to the wire (blue). The orientation of $B$ is visualized by the insets. (b-c), Color map of the bin counts of sample $r1$ at $B_\perp = 10$ mT for $f = 5.9$ GHz and $f = 5.4$ GHz. The magnetic field is aligned perpendicular to the wire. In (b) the first Shapiro step is reduced, while it is fully absent in (c). (d-e) Shapiro maps of sample $r1$ at $B_\perp = 30$ mT for $f = 3.7$ GHz and $f = 3.1$ GHz. While in (d) the first step is still visible, in (e) it is strongly reduced. (f) The $4\pi$-periodic current $I_{4\pi}$ as a function of the flux (calculated from the parallel magnetic field data) for sample r1 estimated by the RSJ-model. Our results suggest that the $4\pi$-periodic current $I_{4\pi}$ in the range up to ~30 mT stems from trivial effects like Landau-Zener transitions, while for higher fields an additional – likely topological – $4\pi$-periodic component arises.